\documentstyle[epsfig]{mn}

\begin{document}
\def\ltsima{$\; \buildrel < \over \sim \;$}
\def\simlt{\lower.5ex\hbox{\ltsima}}
\def\gtsima{$\; \buildrel > \over \sim \;$}
\def\simgt{\lower.5ex\hbox{\gtsima}}

\title[The iron K$\alpha$ Compton Shoulder in transmitted and 
reflected spectra]
{The iron K$\alpha$ Compton Shoulder in transmitted and
reflected spectra}

\author[Giorgio Matt]
{Giorgio Matt \\ ~ \\
Dipartimento di Fisica, Universit\'a degli Studi ``Roma 
Tre'', Via della Vasca Navale 84, I-00146 Roma, Italy \\
}

\maketitle
\begin{abstract}
We calculate the Equivalent Widht of the Core and the centroid energy and relative
flux of the 1st order Compton Shoulder of the iron K$\alpha$ emission line from
neutral matter. The calculations are performed with Monte Carlo simulations.
We explore a large range of column densities for both transmitted
and reflected spectra, and study the dependence on the iron abundance. The 
Compton Shoulder is now becoming observable in many objects
thanks to the improved sensitivity and/or
energy resolution of XMM--$Newton$ and $Chandra$ satellites, 
and the present work aims to provide
a tool to derive informations on the geometry and element abundances
of the line emitting matter from Compton Shoulder measurements. 
\end{abstract}


\section{Introduction}

Iron K$\alpha$ fluorescent lines emitted in neutral matter consists
(in the matter rest frame) of a narrow core (corresponding to the line
photons emerging unscattered from the emitting region) and several Compton
Shoulders (CS; see Matt et al. 1991; George \& Fabian 1991; Leahy \&
Creighton 1993; Sunyaev \& Churazov 1996), corresponding to line photons emerging 
after one or more scatterings. While the higher order CS are expected
to be very faint, the first order CS (hereinafter CS1)
should now be observable in many objects thanks to 
the improved energy resolution and sensitivity of the instruments
onboard $Chandra$ and XMM--Newton (see Kaspi et al. 2002; Bianchi et al. 2002; see also
Iwasawa et al. 1997 for the only pre-$Chandra$ observation of a Compton Shoulder). 

In this paper we calculate, by means of Monte Carlo simulations, the 
relative amount and centroid energy  of CS1, as well as the
Equivalent Width (EW) of the unscattered line photons (Narrow Core, 
hereinafter NC),  in both transmitted and reflected spectra. In the former case, 
we adopt a spherical geometry, and explore a large interval of the column density.
For the latter case we assume a plane--parallel
geometry and calculate CS1 properties 
as a function of the inclination angle for different values of the 
column density in the perpendicular direction. The dependence of line properties
on the iron abundance is also explored.

\section{Calculations}

The simulation code employed here is described in Matt et al. (1991; 1997),
and we defer the reader to these papers for details. To avoid contamination
from Compton scattered continuum photons, we injected only primary photons
with energies larger than the iron K$\alpha$ edge.

For each simulated spectrum, we calculated three quantities:
a) the EW of the NC with respect to the primary (unabsorbed) emission;
b) the Centroid Energy ($E_c$) of CS1, defined as:

\begin{equation}
E_c = { \int^{6.4}_{E_{\rm rec}} E N(E) dE \over  
\int^{6.4}_{E_{\rm rec}} N(E) dE }
\end{equation}

\noindent
where $E_{\rm rec}$=6.2436 keV is the mimimum energy of a line photon after one 
scattering, $N(E)$ is the number of line photons per unit energy at a given
energy, and the integral does not of course include the NC; c) the ratio
$f$ between the total number of photons in the CS1 
($=\int^{6.4}_{E_{\rm rec}} N(E) dE$) to that in the NC. It should be noted
that with these definitions, the CS1 may actually include also line photons 
scattered twice or more times, provided that the energy loss per 
scattering is small enough (i.e. small scattering angles) that the emerging photon
has an energy still larger than  $E_{\rm rec}$. However, 
the chosen definitions are the most useful when comparing the calculations
with real data.

Element abundances and photoabsorption cross
sections are those given by Morrison \& McCammon (1983) (the iron abundance
in number is  3.3$\times10^{-5}$; a value about 40 percent higher is
given by Anders \& Grevesse 1989). A larger (smaller)
iron abundance results in an increase (decrease) of the EW
of the iron line (see Matt et al. 1997 for a detailed discussion),
and in a decrease (increase) of $f$
due to the larger (smaller) ratio of photoelectric absorption 
to Compton scattering. To quantify these effects, we have run the
Montecarlo code (in the optically thick case only for 
reflected spectra) also for $A_{Fe}$=0.5 
and 2, where $A_{Fe}$ is the iron abundance in units of the cosmic value.

The primary spectrum has been assumed to be a power law with photon
spectral index 2. Different indices will result on somewhat
different values of the EW of the NC (see George \& Fabian 1991). $E_c$ and 
$f$ are instead independent of the spectral index, at least in the optically
thin case (in the optically thick case there should be a small dependence
on the slope due to the energy dependence of photoabsorption, and therefore
of the depth at which the interaction occurs.)

Finally, it must be remarked that we assumed cold electrons, i.e. we neglected
both thermal motions for free electrons and their motion in the
atoms and molecules, if they are bound. In real situations, where these effects are
considered, the minimum
energy after one scattering, $E_{\rm rec}$, may be lower than the value given above. 
The adopted approximation will also lead to an artificially
sharp backscattering peak in the spectrum of Figs. 2 and 5 below, and this should
be considered when fitting real data. (For a complete
discussion of these effects, see Sunyaev \& Churazov, 1996.) We note that
electron temperatures are expected to be large in the inner regions of 
accretion discs, expecially in Galactic Black Hole candidates, but then
relativistic distortion are expected to be even larger. Electron temperatures
are instead expected to be low in e.g. circumnuclear matter in AGN; in these
cases, effects due to the motion of bound electrons should instead be considered
(see e.g. Fig.~4a of  Sunyaev \& Churazov 1996).

\section{Results}

\subsection{CS1 in transmitted spectra}

We first calculated the line spectra when the cold matter lies along
the line of sight (transmitted spectra). We explored a wide
range of column densities, from 10$^{22}$ to 5$\times10^{24}$ cm$^{-2}$.
We assumed a spherical distribution of the matter, with the source of the primary 
photons located at the centre of the sphere. 
The density of the matter is assumed to be constant.

\begin{figure}
\epsfig{figure=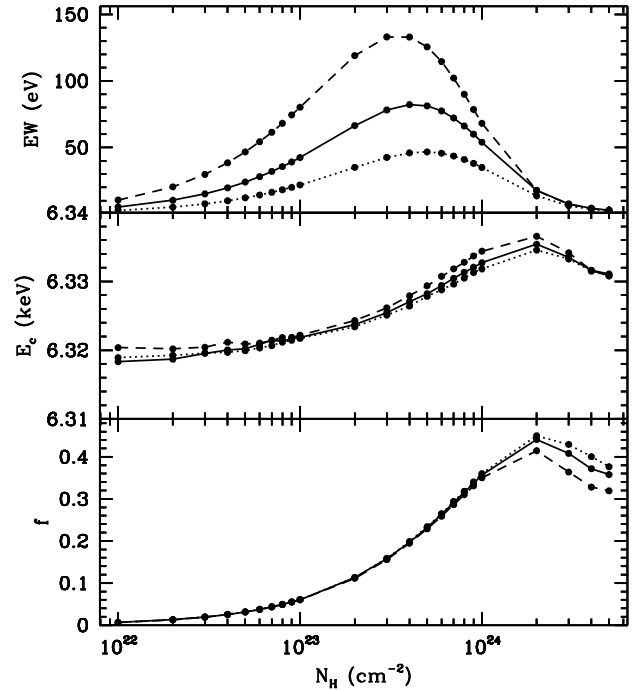,height=10.0cm,width=10.0cm}
\caption{The EW of the NC (calculated with respect to the primary unabsorbed
emission), centroid energy of CS1 and $f$ for a centrally
illuminated spherical distribution of cold matter. The dotted, solid and dashed lines
refer to $A_{Fe}$=0.5, 1 and 2, respectively.}
\label{sph}
\end{figure}

We explored the case of spherical distribution of matter for
three different values of the iron abundance $A_{Fe}$ with respect to the
cosmic value (as for Morrison \& McCammon 1983): 0.5 (Fig.~\ref{sph}, dotted lines), 1
(solid lines), 2 (dashed lines). The EW of the NC increases
with the column density up to about 4$\times10^{23}$ cm$^{-2}$, in agreement
with previous calculations (e.g. Inoue 1989), and then decreases. The dependence
on the iron abundance is linear in the $\tau_C\ll$1 limit 
($\tau_C=N_H/1.5\times10^{24}$~cm$^{-2}$), 
and less than linear when opacity effects become important (see Matt et al. 1997). 
The CS1
centroid energy and $f$  increase up to a Compton depth of about unity.
The following decrease is due to the increasing importance of repeated scatterings.
The dependence of both parameters on $A_{Fe}$ is not dramatic and, for $f$,
almost negligible up to $\tau_C\sim$1.

\begin{figure}
\epsfig{figure=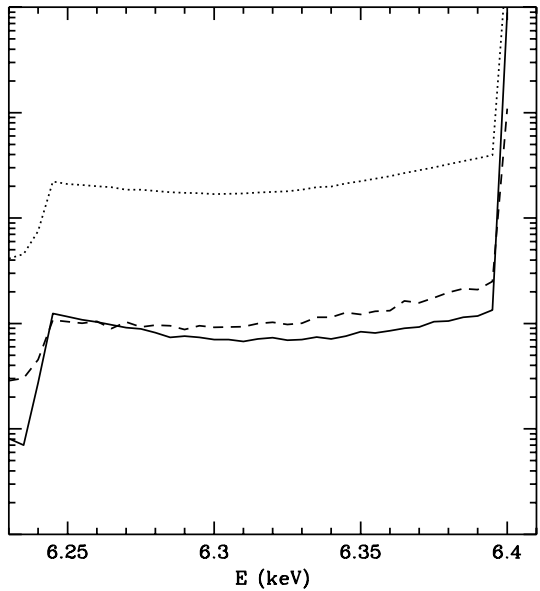,height=10.0cm,width=10.0cm}
\caption{The spectrum of CS1 (in the cold electrons approximation, see text)
in the transmitted case for column densities 
of 5$\times10^{22}$ (solid line),  5$\times10^{23}$ (dotted line) and
 5$\times10^{24}$ cm$^{-2}$ (dashed line). $A_{Fe}$=1.}
\label{sp_tr}
\end{figure}

In Fig.~\ref{sp_tr} the spectrum of the CS1 is shown for three different column densities
($A_{Fe}$=1).

\subsection{CS1 in reflected spectra}

\begin{figure}
\epsfig{figure=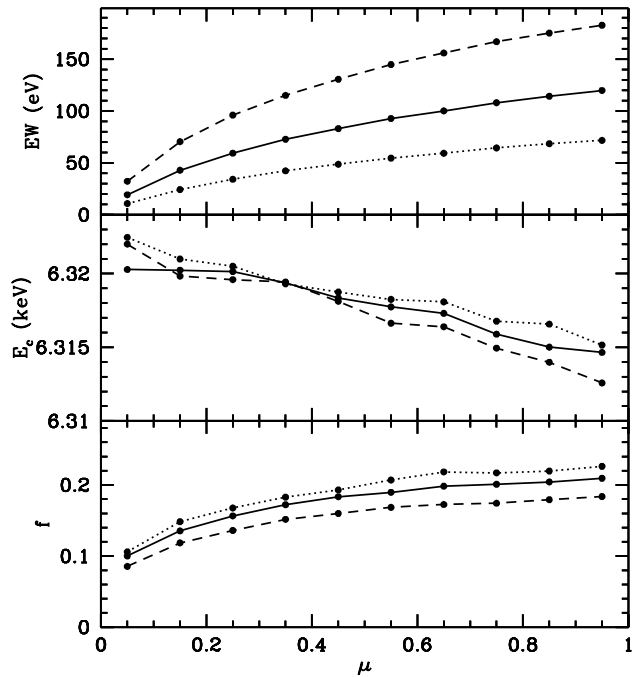,height=10.0cm,width=10.0cm}
\caption{The EW of the NC, centroid energy of CS1 and $f$ for an isotropically
illuminated, optically thick ($N_H=2\times10^{25}$ cm$^{-2}$)
plane--parallel slab, as a function of $\mu$, the cosine of the inclination angle. 
The dotted, solid and dashed 
lines refer to $A_{Fe}$=0.5, 1 and 2, respectively.}
\label{refl}
\end{figure}

We then calculated the CS1 properties for an isotropically illuminated plane--parallel slab. 
The EW of the NC (with respect to the primary emission only), the  centroid energy of CS1 
and $f$ are shown in Fig.~\ref{refl} (solid lines) as a function of $\mu$, the cosine
of the slab's inclination angle. We firstly assumed optically thick matter
(vertical column density of 2$\times10^{25}$ cm$^{-2}$) and cosmic abundances. 
The EW of the NC increases with $\mu$ 
(as well known, see Matt et al. 1991; George \& Fabian 1991), as does $f$.
$E_c$, instead, decreases with $\mu$.

Is is worth noting that in reflected spectra
$f$, for the same iron abundance, is lower than in trasmitted spectra, provided that the column
density of the line--of--sight matter is larger than a few $\times10^{23}$ cm$^{-2}$.
 This is because the
average optical depth, with respect to the emerging surface, 
at which the first interaction occurs is lower in the former than in the
latter case due to the isotropic illumination.

We then explored the dependence of the line properties 
on the iron abundance. In Fig.~\ref{refl} 
the results for A$_{Fe}$=0.5 and 2 are shown (dotted and dashed lines, 
respectively). As expected,
the EW is larger and $f$ smaller for the larger iron abundance, while the opposite
occurs for the smaller iron abundance. 

\begin{figure}
\epsfig{figure=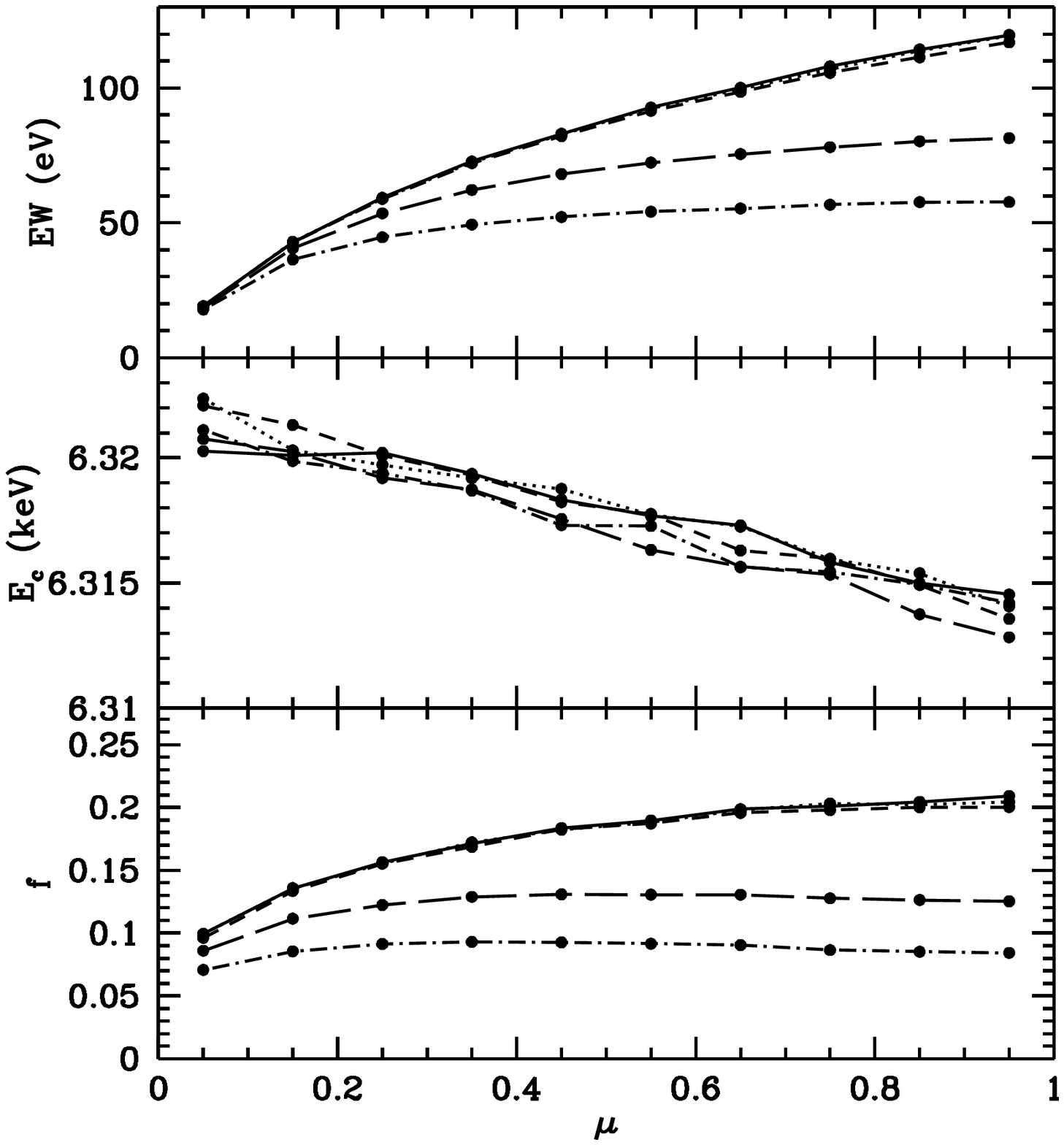,height=10.0cm,width=10.0cm}
\caption{The EW of the NC, centroid energy of CS1 and $f$ for an isotropically
illuminated, plane--parallel slab with vertical column density of 
$5\times10^{22}$ (dot--dashed lines),
$10^{23}$ (long--dashed lines), $5\times10^{23}$ (short--dashed lines), 
$10^{24}$ (dotted lines) and
$5\times10^{24}$ (solid lines) cm$^{-2}$. $A_{Fe}$=1.}
\label{refl_thin_all}
\end{figure}


Finally, we calculated the line properties 
as a function of the vertical column density (Fig.~\ref{refl_thin_all}).
While the centroid energy is not significantly dependent on $N_H$, both $f$ and the
EW of the core increases, as expected, with $N_H$ in the optically thin regime, to saturate
for $N_H\sim5\times$10$^{23}$ cm$^{-2}$. In the Compton--thin regime, moreover,
these parameters are much less dependent on $\mu$ than in the optically thick case.
It is worth noting that in the optically thin regime the values of EW and $f$ are larger 
than in transmitted spectra, for the same $N_H$, largely due to the different 
illumination (which is always normal to the surface in the spherical geometry, but not
in the isotropically illuminated slab; in the latter case the effective column density 
becomes very large for grazing angles). 

\begin{figure}
\epsfig{figure=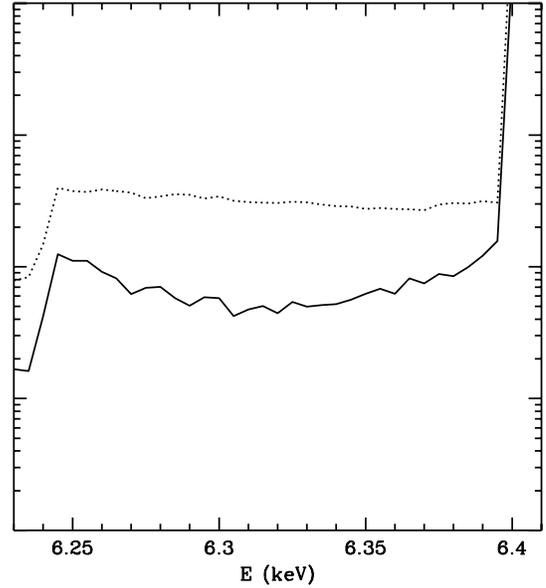,height=10.0cm,width=10.0cm}
\caption{The spectrum of CS1 (in the cold electrons approximation, see text)
in the reflection case for $\mu$=0.15 
(solid line) and $\mu$=0.95 (dotted line). $A_{Fe}$=1.}
\label{sp_ref}
\end{figure}

In Fig.~\ref{sp_ref} the spectrum of CS1 is shown for two different values of $\mu$
($A_{Fe}$=1).

\section{Conclusions}

We have calculated the properties of the iron K$\alpha$
Narrow Cores and of the first scattering Compton Shoulder in both 
transmitted and reflected spectra, refining and expanding previous works on the same subject
(Matt et al. 1991; George \& Fabian 1991; Leahy et al. 1993; Iwasawa et al. 1997;
see Sunyaev \& Churazov 1996 for a very detailed physical description of the process). 
The intensity of the Compton Shoulder is at most 30-40\% of the Narrow
Core of the line, with a centroid energy typically of about 6.3 keV or
more. 

Prior to the launch of $Chandra$, the Compton Shoulder was observed only in the
ASCA spectrum of NGC~1068 (Iwasawa et al. 1997), with properties consistent
with an origin in the Compton--thick torus of this source. Recently, thanks to
the gratings onboard $Chandra$, two more cases have been reported. Bianchi et
al. (2002) found a Compton shoulder in the HETG spectrum 
of the Circinus Galaxy. The value of $f$, i.e. about
20\%, is consistent with reflection from the inner wall of the 4$\times10^{24}$
cm$^{-2}$ torus.

Kaspi et al. (2002) measured a Compton shoulder in their
900 ks $Chandra$/HETG observation of the Seyfert 1 galaxy NGC~3783. The flux
in CS1 is 14$\pm$4 percent of the narrow core, which in turn has an EW of 90$\pm$11 eV.
The line is clearly due to reflecting matter, as no evidence for strong cold absorption
is apparent. 
Looking at the values of $f$ in Figs.~\ref{refl} and \ref{refl_thin_all},
it is possible to conclude that the reflecting matter should have a column
density of at least $\sim10^{23}$ cm$^{-2}$ (slightly lower values are
allowed only if the iron is overabundant).
The value of the EW is less directly usable, as the geometry of the reflector is
unknown (the values in this paper refer to a 2$\pi$ solid angle of
the matter, subtended to an isotropic primary source). 
However, as it seems unlikely that the solid angle is much larger than 2$\pi$,
low values of the iron abundance are disfavoured.

\section*{Acknowledgments}

I thank the referee, E. Churazov, for his useful comments. 
I acknowledge ASI and MIUR (under
grant {\sc cofin-00-02-36}) for financial support.

{}

\end{document}